\documentclass[aps,prl,reprint,superscriptaddress]{revtex4-1}
\usepackage{amsmath}
\usepackage{amssymb}
\usepackage{graphicx}

\newcommand{\Ciso}{$^{13}$C}
\newcommand{\Cis}{$^{12}$C}
\newcommand{\Niso}{$^{14}$N}
\newcommand{\Nis}{$^{15}$N}
\newcommand{\NVz}{\ensuremath{\text{NV}^{0}}}
\newcommand{\NVm}{\ensuremath{\text{NV}^{-}}}
\newcommand{\NVp}{\ensuremath{\text{NV}^{+}}}

\newcommand{\gammaered}{\ensuremath{\tilde{\gamma}_\mathrm{e}}}
\newcommand{\gammanred}{\ensuremath{\tilde{\gamma}_\mathrm{n}}}

\newcommand{\ket}[1]{\ensuremath{\left| #1 \right \rangle}}

\begin{document}

%\preprint{}

%\title{Protection of a nuclear spin storage qubit by switching off the electron spin bus qubit.}
\title{Protecting a diamond quantum memory by charge state control}

\author{Matthias Pfender}
\author{Nabeel Aslam}
\affiliation{3rd institute of physics, University of Stuttgart, Pfaffenwaldring 57, 70569 Stuttgart, Germany}
\affiliation{Stuttgart Research Center of Photonic Engineering (SCoPE), Stuttgart, Germany}
\affiliation{Center for Integrated Quantum Science and Technology (IQST), Stuttgart, Germany}

\author{Patrick Simon}
\affiliation{Walter Schottky Institut, Physik-Department, Technische Universit\"at M\"unchen, Am Coulombwall 3, 85748 Garching, Germany}

\author{Denis Antonov}
\affiliation{3rd institute of physics, University of Stuttgart, Pfaffenwaldring 57, 70569 Stuttgart, Germany}
\affiliation{Stuttgart Research Center of Photonic Engineering (SCoPE), Stuttgart, Germany}
\affiliation{Center for Integrated Quantum Science and Technology (IQST), Stuttgart, Germany}

\author{Gerg\H{o} Thiering}
\affiliation{Institute for Solid State Physics and Optics, Wigner Research Centre for Physics, Hungarian Academy of Sciences, Budapest, PO Box 49, H-1525, Hungary}
\affiliation{Department of Atomic Physics, Budapest University of Technology and Economics, Budafoki \'ut 8, H-1111 Budapest, Hungary}

\author{Sina Burk}
\author{Felipe F\'avaro de Oliveira}
\author{Andrej Denisenko}
\author{Helmut Fedder}
\affiliation{3rd institute of physics, University of Stuttgart, Pfaffenwaldring 57, 70569 Stuttgart, Germany}
\affiliation{Stuttgart Research Center of Photonic Engineering (SCoPE), Stuttgart, Germany}
\affiliation{Center for Integrated Quantum Science and Technology (IQST), Stuttgart, Germany}

\author{Jan Meijer}
\affiliation{Institute for Experimental Physics II, Universit\"{a}t Leipzig, Linn\'{e}stra\ss e 5, 04103 Leipzig, Germany}

\author{Jose Antonio Garrido}
\affiliation{Catalan Institute of Nanoscience and Nanotechnology (ICN2), CSIC, Barcelona Institute of Science and Technology, Campus UAB, Bellaterra, Barcelona, Spain}
\affiliation{Walter Schottky Institut, Physik-Department, Technische Universit\"at M\"unchen, Am Coulombwall 3, 85748 Garching, Germany}

\author{Adam Gali}
\affiliation{Institute for Solid State Physics and Optics, Wigner Research Centre for Physics, Hungarian Academy of Sciences, Budapest, PO Box 49, H-1525, Hungary}
\affiliation{Department of Atomic Physics, Budapest University of Technology and Economics, Budafoki \'ut 8, H-1111 Budapest, Hungary}

\author{Tokuyuki Teraji}
\affiliation{National Institute for Materials Science, 1-1 Namiki, Tsukuba, Ibaraki 305-0044, Japan}
\author{Junichi Isoya}
\affiliation{Research Center for Knowledge Communities, University of Tsukuba, Tsukuba, 305-8550 Japan}

\author{Marcus W. Doherty}
\affiliation{Laser Physics Centre, Research School of Physics and Engineering, Australian National 	University, Australian Capital Territory 0200, Australia}

\author{Audrius Alkauskas}
\affiliation{Center for Physical Sciences and Technology, Vilnius LT-10257, Lithuania}

\author{Alejandro Gallo}
\affiliation{Max Planck Institute for Solid State Research, Stuttgart, Germany}
\affiliation{3rd institute of physics, University of Stuttgart, Pfaffenwaldring 57, 70569 Stuttgart, Germany}

\author{Andreas Gr\"uneis}
%\affiliation{Computergest\"utzte Quantenchemie f\"ur Festk\"orper, Max Planck Institute for Solid State Research, Stuttgart, Germany}
\affiliation{Max Planck Institute for Solid State Research, Stuttgart, Germany}

\author{Philipp Neumann}
\email[]{p.neumann@physik.uni-stuttgart.de}
\affiliation{3rd institute of physics, University of Stuttgart, Pfaffenwaldring 57, 70569 Stuttgart, Germany}
\affiliation{Stuttgart Research Center of Photonic Engineering (SCoPE), Stuttgart, Germany}
\affiliation{Center for Integrated Quantum Science and Technology (IQST), Stuttgart, Germany}

\author{J\"org Wrachtrup}
\affiliation{3rd institute of physics, University of Stuttgart, Pfaffenwaldring 57, 70569 Stuttgart, Germany}
\affiliation{Stuttgart Research Center of Photonic Engineering (SCoPE), Stuttgart, Germany}
\affiliation{Center for Integrated Quantum Science and Technology (IQST), Stuttgart, Germany}
\affiliation{Max Planck Institute for Solid State Research, Stuttgart, Germany}

\date{\today}

\begin{abstract}
In recent years, solid-state spin systems have emerged as promising candidates for quantum information processing (QIP).
Prominent examples are the nitrogen-vacancy (NV) center in diamond \cite{doherty_nitrogen-vacancy_2013,dutt_quantum_2007,waldherr_quantum_2014}, phosphorous dopants in silicon (Si:P) \cite{morton_solid-state_2008,saeedi_room-temperature_2013,pla_high-fidelity_2013}, rare-earth ions in solids \cite{kolesov_mapping_2013,siyushev_coherent_2014,kolesov_optical_2012} and $\mathrm{V_{Si}}$-centers in silicon-carbide (SiC) \cite{christle_isolated_2015,koehl_room_2011,widmann_coherent_2015}.
The Si:P system has demonstrated that its nuclear spins can yield exceedingly long spin coherence times by eliminating the electron spin of the dopant.
For NV centers, however, a proper charge state for storage of nuclear spin qubit coherence has not been identified yet \cite{waldherr_dark_2011}.
Here, we identify and characterize the positively-charged NV center as an electron-spin-less and optically inactive state by utilizing the nuclear spin qubit as a probe \cite{waldherr_dark_2011}.
We control the electronic charge and spin utilizing nanometer scale gate electrodes.
We achieve a lengthening of the nuclear spin coherence times by a factor of 20.
Surprisingly, the new charge state allows switching the optical response of single nodes facilitating full individual addressability.
\end{abstract}


%\keywords{}

\maketitle

\begin{figure}[t]
%\begin{figure*}[t]
	\includegraphics[width=1.0\columnwidth]{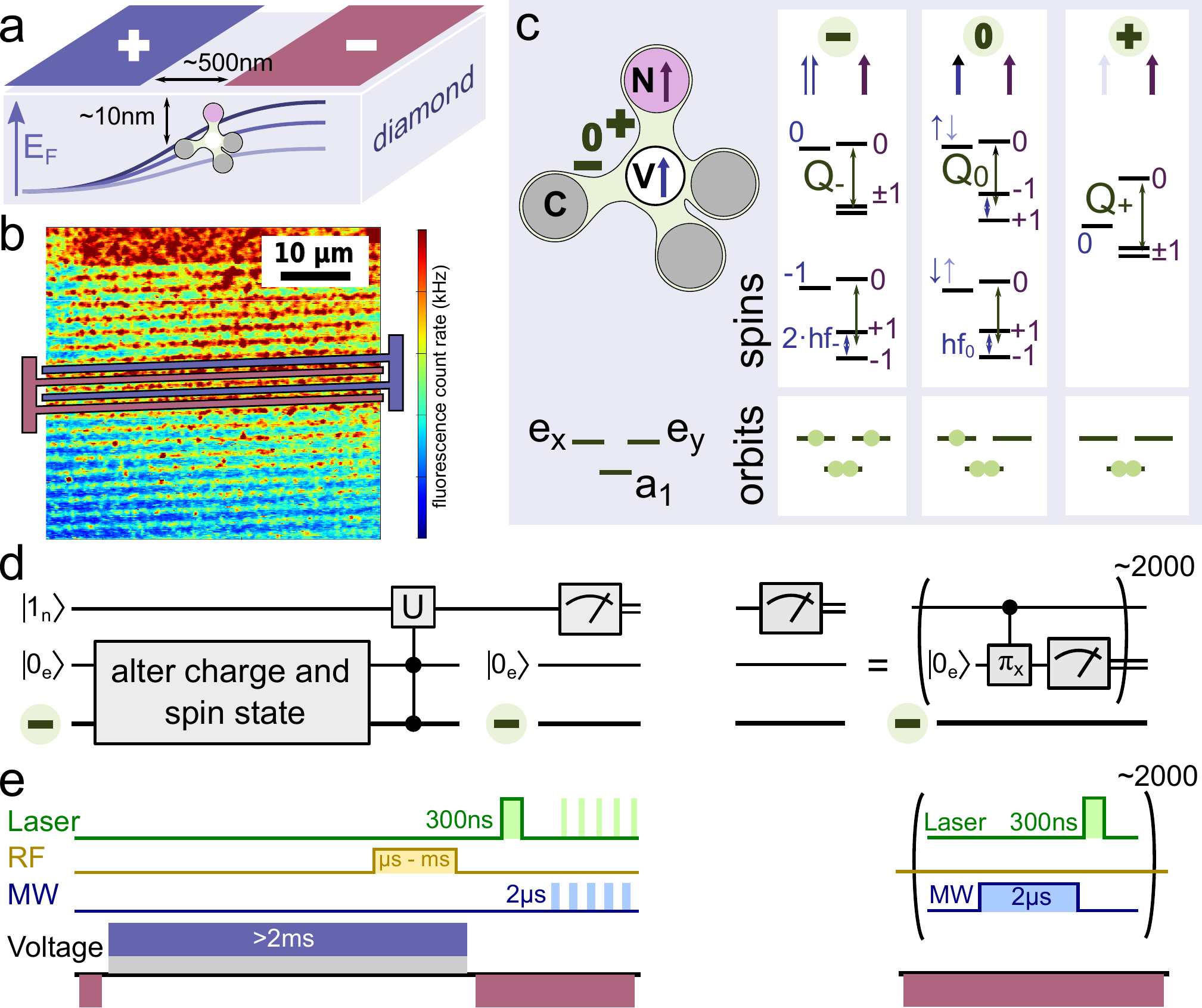}
	\caption{\label{fig:charge_state_switching}
		\textbf{NV center in diamond and its charge states.}
		%\todo{Each legend should total no more than 250 words.}
		(\textbf{a}) Principle of gate voltage induced charge state switching.
		The voltage difference between the two hydrogen terminated (hence conductive) stripes of the diamond surface (blue +, red $-$), affects the Fermi level at the location of the NV center under one H-terminated stripe and close to an oxygen-terminated (non-conductive) diamond stripe (pale blue).
		If the Fermi level crosses the charge transition level, a change in the dominant charge state is observed \cite{hauf_addressing_2014}.
		(\textbf{b}) Confocal microscopy scan of the diamond surface revealing the gate structure.
		Overlayed schematics illustrates parts of the two hydrogen terminated electrodes in form of an interdigitated capacitor.
		(\textbf{c}) Sketch of the NV center in diamond containing the nitrogen nuclear spin (purple arrow), the electron spin (blue arrow) and the electronic wave function (green).
		The right side shows the electronic occupation of the orbitals $e_x,\,e_y$ and $a_1$ within the diamond bandgap \cite{doherty_nitrogen-vacancy_2013} for three different charge states (``+'', ``0'', ``$-$'').
		Only the unpaired electrons in the $e_{x,y}$ orbitals contribute to the electron spin.
		The corresponding electron and nuclear spin energy levels are sketched above for the case of \Niso.
		For the \Nis\ case the $m_I=0$ level and hence the quadrupole splitting is obsolete, $m_I=\pm1$ become $\pm1/2$ and the hyperfine and Zeeman terms change.
		(\textbf{d}) wire diagram of the general measurement scheme.
		The main workhorse is the charge and spin state selective quantum gate $U$ on the nitrogen nuclear spin.
		(\textbf{e}) shows the corresponding physical implementation utilizing a sequence of Laser, microwave (MW) and radiofrequency (RF) fields and the applied voltage.
  }
\end{figure}
\begin{figure}
	\includegraphics[width=1.0\columnwidth]{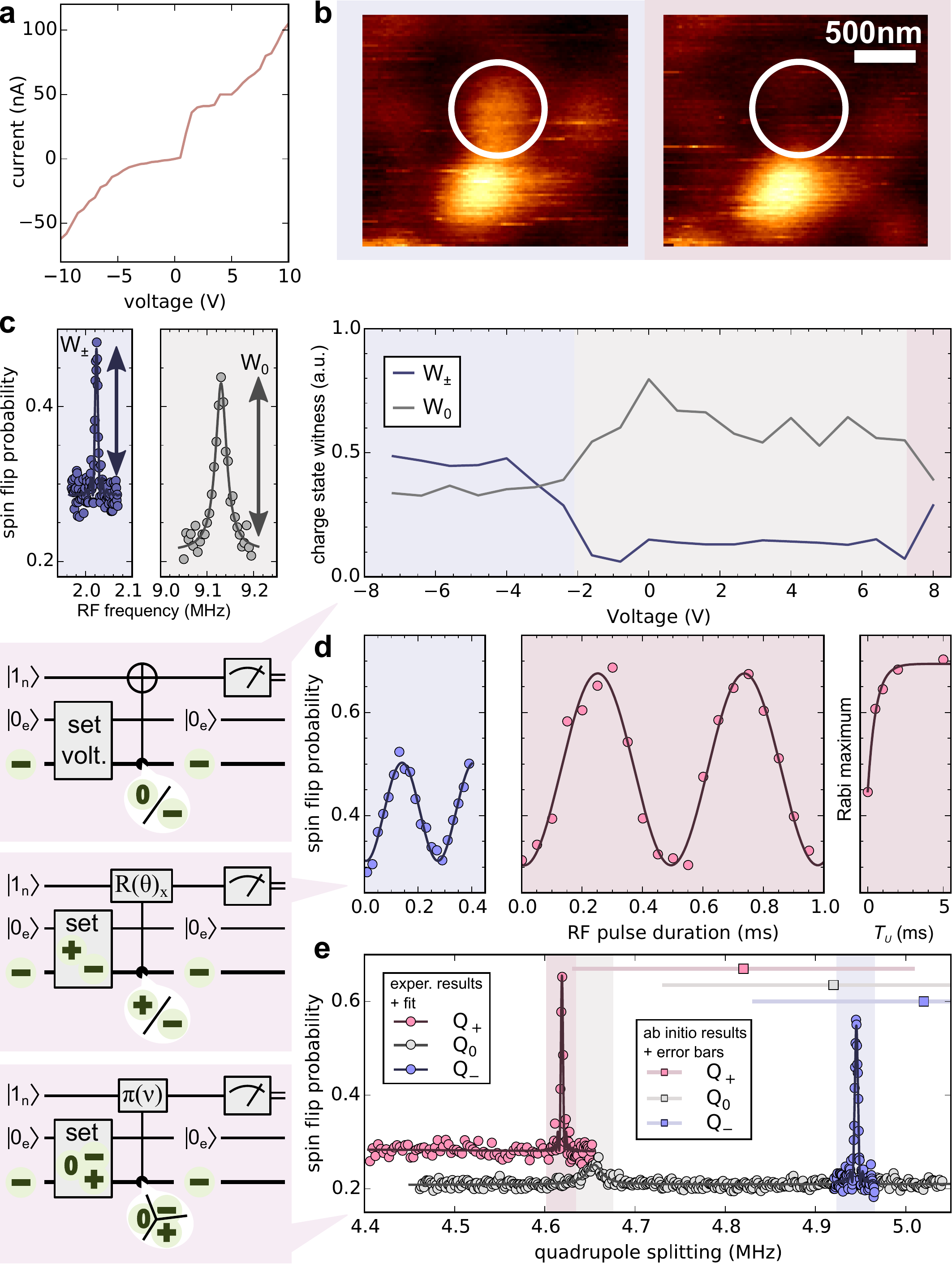}
	\caption{\label{fig:NVplus}
		\textbf{Characterizing the positively charged NV center.}
		(\textbf{a}) shows the I-V-characteristic of the surface gate structure.
		(\textbf{b}), confocal microscopy image of single NV centers.
		The left image (light blue background) was recorded at the voltage corresponding to the \NVm\ charge state.
		In the right image (light red background) the NV's fluorescence is quenched by reversing the gate voltage.
		The background colors denote the same charge state throughout the figure.
		(\textbf{c}) The two spectra on the left show one of the \Nis\ nuclear spin transitions for the \NVm\ and the \NVz\ charge state, respectively.
		Their normalized amplitudes yield the charge state probabilities $W_\pm$ and $W_0$.
		$W_\pm$ would increase for any charge state exhibiting an electron spin state $m_S=0$, as is expected for \NVp\ ($S=0,\,m_S=0$).
		The graph on the right shows the charge state probabilities for varying voltages.
		The reappearance of the $W_\pm$ signal at high voltages suggests the presence of \NVp .
		(\textbf{d}) \Nis\ nuclear spins Rabi oscillations between spin state $m_I=\pm 1/2$ for the \NVm\ ($m_S=0$) and the tentative \NVp\ charge state, respectively.
		The right curve shows the maxima of the \NVp\ Rabi oscillation for varying duration $T_U$ of initial \NVp -voltage application.
		We deduce an \NVp\ settling time of $0.54\pm0.08\,$ms.
		For the \NVm\ charge state, hyperfine interaction increases the Rabi frequency (see text and Methods).
		%The Rabi frequency in the \NVm\ charge state is increased by dressing of the nuclear spin states by the electron spin $S=1$ by a factor of $1.81\pm0.04$.
%		The smaller Rabi frequency for the \NVp\ case is in very good agreement with a bare \Nis\ nuclear spin ($1.832$, see Methods section), without any electron spin.
		(\textbf{e}), \Niso\ quadrupole splittings for the three known charge states.
		Any hyperfine and nuclear Zeeman terms are subtracted.
  }
\end{figure}
\begin{figure}
	\includegraphics[width=1.0\columnwidth]{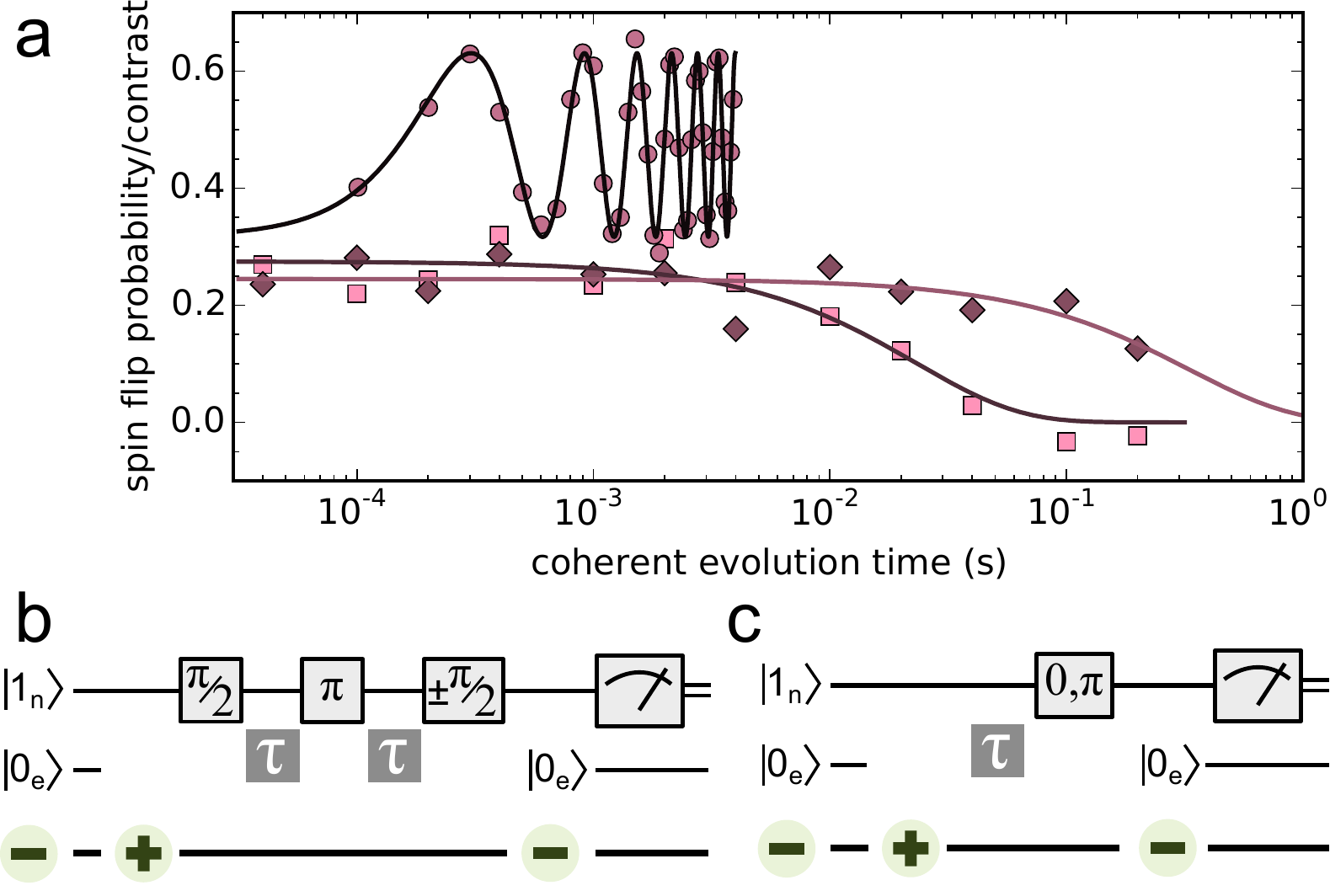}
	\caption{\label{fig:long_lifetime}
		\textbf{Protected nuclear spin quantum memory.}
		(\textbf{a}) Increased nuclear spin lifetimes in the \NVp\ charge state.
		The spin-echo coherence time (squares) is $25\pm10\,$ms and the longitudinal relaxation time (diamonds) is $0.3\pm1.4\,$s.
		We expect both values to be limited by paramagnetic noise originating from the surface and sub-surface impurities.
		The Rabi oscillation (circles) has a decay constant of $22\pm12\,$ms.
		(\textbf{b}) and (\textbf{c}) show the wire diagram for the spin-echo and the $T_1$ measurement, respectively.
  }
\end{figure}
\begin{figure}
	\includegraphics[width=1.0\columnwidth]{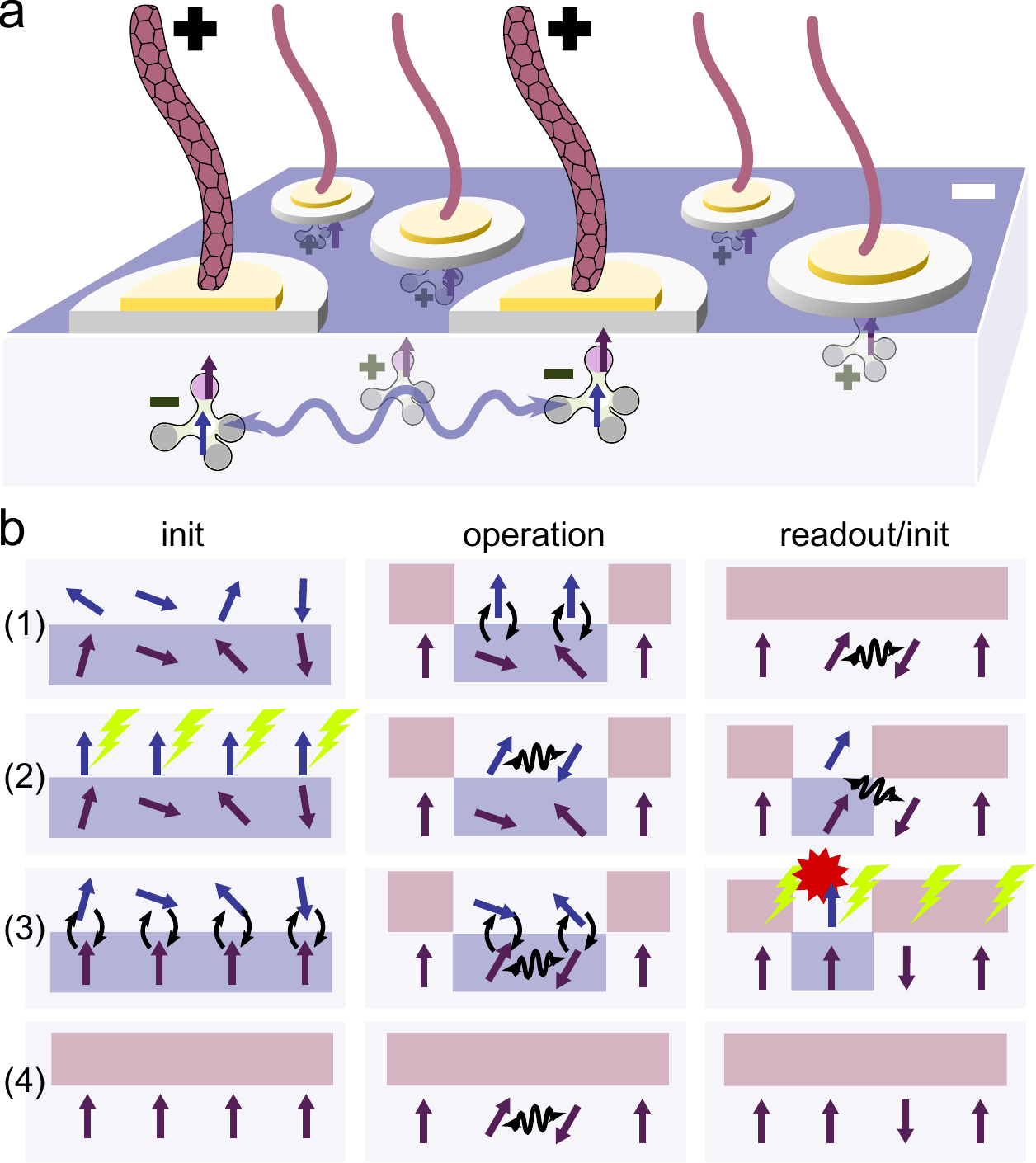}
	\caption{\label{fig:scalable}
		\textbf{Proposal for a scalable NV diamond quantum processor.}
		(\textbf{a}) Schematic of individual NV nodes addressable via nanoscopic gate electrodes (gray: insulator, gold: electrode, purple: leads).
		NV centers under H-terminated surface are in their \NVp\ state and therefore do not couple to other nodes because of the lack of an electron spin, their nuclear spin state remains undisturbed and they do not contribute to fluorescence response when illuminated.
		Individual electrodes with a positive voltage deplete the hole conducting layer locally and shift the Fermi level into the \NVm-stable region.
		Therefore, such NV centers can be coupled to their \NVm\ neighbors via magnetic dipole interaction and they are optically accessible.
		Unsuitable NV nodes are not supplied with an electrode and remain in \NVp .
		(\textbf{b})  Sketch of quantum register operation modes using red and blue boxes to highlight charge states, purple and blue arrows for nuclear and electron spin, green lightnings for laser and red stars for fluorescence.
		\textbf{Initialization:} (1)-(2) Laser initializes all electron spins, (3) swap to nuclear spins and (4) switch to \NVp\ for storage.
		\textbf{Operation:} (1) Swap two nuclear spins to electron spins in \NVm , (2) entangle electron spins, (3) swap back to nuclei and (4) switch all NV centers to \NVp .
		\textbf{readout/init:} (1) Switch one NV to \NVm with initialized electron spin, (2) correlate electron spin with nuclear spin, (3) readout one electron spin and project nuclear spins (2-3, single shot readout). (4) Switch back to \NVp .
  }
\end{figure}

Spin defects are excellent quantum systems.
Particularly, defects that possess an electron spin together with a set of well-defined nuclear spins make up excellent, small quantum registers \cite{pla_coherent_2014,waldherr_quantum_2014}.
They have been used for demonstrations in quantum information processing \cite{dolde_room-temperature_2013,waldherr_quantum_2014}, long distance entanglement \cite{bernien_heralded_2013} and sensing \cite{zaiser_enhancing_2016}.
In such systems the electron spin is used for efficient readout (sensing or interaction with photons) whereas the nuclear spins are used as local quantum bits.
Owing to their different magnetic and orbital angular momentum, electron and nuclear spins exhibit orders of magnitude different spin relaxation times.
As an example, NV electron spins typically relax on a timescale of ms under ambient conditions \cite{jarmola_temperature-_2012}, while nuclear spins do have at least minutes-long spin relaxation times \cite{pfender_nonvolatile_2016}.
However, the hyperfine coupling of nuclear spins to the fast relaxing electron spins in most cases significantly deteriorates the nuclear spin coherence, and eventually its relaxation time, down to time scales similar to the electron spin.
For NV centers at room-temperature, this limits coherence times to about $10\,$ms \cite{waldherr_high-dynamic-range_2011,zaiser_enhancing_2016,pfender_nonvolatile_2016}.
For other hybrid spin ensembles e.g. in Si:P, this strict limit was overcome by ionizing the electron spin donors and thereby removing the electron spin. % after nuclear spin initialization.
The resulting $T_2$ times were on the order of minutes for Si:P ensembles \cite{saeedi_room-temperature_2013} and less than a second for single spins \cite{morello_single-shot_2010,pla_high-fidelity_2013}.
\\
\indent
It is known that the NV center in diamond exists in various charge states.
Besides the widely employed negative charge state (\NVm), it is known to have a stable \NVz\ and eventually \NVp\ configurations.
\NVm\ has a spin triplet ground state with total spin angular momentum $S=1$.
Calculations as well as spectroscopic data suggest that the \NVz\ ground state is $S=1/2$ while \NVp\ is believed to be $S=0$, i.e. diamagnetic.
Several experiments have demonstrated the optical ionization from \NVm\ to the neutral \NVz\ charge-state \cite{han_metastable_2010,waldherr_dark_2011,han_dark_2012,doi_deterministic_2014,aslam_photo-induced_2013,beha_optimum_2012} and electroluminescence from single \NVz\ centers \cite{mizuochi_electrically_2012}.
These experiments characterize the \NVm\ and \NVz\ charge states via their photophysics and different fluorescence spectral fingerprints (see Fig.~\ref{fig:charge_state_switching}c).
\\
\indent
Apart from optical ionization, deterministic electronic charge state control is feasible using a hydrogen-terminated (H) diamond surface with in-plane or electrolyte gates for  Fermi-level manipulation \cite{hauf_chemical_2011,hauf_addressing_2014,grotz_charge_2012} (cf. Fig.~\ref{fig:charge_state_switching}a).
%Furthermore, the nitrogen nuclear spin qubit of the NV center has been used as a probe to reveal information about the \NVz\ spin system \cite{waldherr_dark_2011} (cf. Fig.~\ref{fig:charge_state_switching}d).
%It was confirmed that the \NVz\ exhibits an electron spin $S=1/2$ in its ground state, which leads to faster nitrogen nuclear spin decoherence than in the \NVm\ state (see Fig.~\ref{fig:charge_state_switching}c).
%\\
%\indent
%Recently, it was shown that gate electrodes can be used to completely quench an NV center's photoluminescence, which was tentatively attributed to switching into a positively-charged state (\NVp) \cite{hauf_addressing_2014} (see Fig.~\ref{fig:charge_state_switching}c).
Hydrogen termination of diamond creates a two-dimensional profile of free holes on the surface.
Recently, it was shown that this surface conductive layer completely quenches the photoluminescence of NV centers in its close proximity.
This optically inactive NV state was tentatively attributed to the positively charged \NVp\ \cite{hauf_addressing_2014} (see Fig.~\ref{fig:charge_state_switching}c).
Proper bias applied to lateral gate electrodes on H-terminated surface depletes the charge carrier density around the gate contact area.
At full depletion, the Fermi level at the surface is shifted away from the valence band edge.
This eventually leads to switching of the charge state ($\mathrm{\NVp \rightarrow \NVz \rightarrow \NVm}$) of such NV centers. 
\\
\indent
Here we show that such nano-scale gate structures can be used to stabilize a third charge state with an $S=0$ ground state under dark conditions, which we attribute to \NVp\ (see Fig.~\ref{fig:charge_state_switching}).
To this end, we apply the intrinsic nitrogen nuclear spin qubit of the NV center as a probe to reveal information about the charge and spin state of the NV center \cite{waldherr_dark_2011} (cf. Fig.~\ref{fig:charge_state_switching}d).
%It was confirmed that the \NVz\ exhibits an electron spin $S=1/2$ in its ground state, which leads to faster nitrogen nuclear spin decoherence than in the \NVm\ state (see Fig.~\ref{fig:charge_state_switching}c).
We utilize this \NVp\ charge state to prolong the nuclear spin $T_2$ lifetime.
Furthermore, the absence of fluorescence allows tailoring the optical response from multiple NV centers within a confocal spot and therefore increases individual addressability.
\\
\indent
%
%
%
%
%\section{Experiment}
To switch the charge state of NV centers we prepare H-terminated, conductive gate electrodes onto an otherwise oxygen-terminated (O) and therefore non-conductive diamond surface \cite{hauf_addressing_2014}.
NV centers have been created approximately 10\,nm below the described surface (see Fig.~\ref{fig:charge_state_switching}b and Methods section).
As H-terminated diamond surfaces quench the fluorescence of NV centers, the gate electrode structure can be seen in the confocal microscopy image (see fig. \ref{fig:charge_state_switching}b).
\\
\indent
In Figure \ref{fig:NVplus}a,b the electrical current and the NV fluorescence response to a varying gate voltage is shown.
For up to $\pm10\,$V the resistance of the capacitor is on the order of $100\,\mathrm{M}\Omega$.
For properly located NV centers the voltage change results in a fluorescence response (see Fig.~\ref{fig:NVplus}b).
The fluorescence of those NV centers located in the center of the H- or O-terminated regions are expected to be
stable under voltage changes, whereas NV centers in the border regions are likely to switch charge states because the Fermi-level change is most pronounced in these regions \cite{hauf_addressing_2014} (cf. Fig.~\ref{fig:charge_state_switching}a).
\\
\indent
%\end{figure*}
%
%
%
%
%\section{charge state nmr}
Next, we concentrate on NV centers that change into a non-fluorescing state upon a suitable voltage change.
For different charge states, the quadrupole splitting $Q$ as well as the hyperfine coupling $A$ of the nitrogen nuclear spin to the NV electron spin vary significantly \cite{waldherr_dark_2011}.
Hence, we can perform charge and electron spin selective control gates on our nuclear spin qubit.
To this end, we initialize the nuclear spin into one of its $m_I$ states ($\ket{1_\mathrm{n}}$ in Fig.~\ref{fig:charge_state_switching}d,e) utilizing single shot readout \cite{waldherr_quantum_2014}.
We set a certain gate voltage resulting in a related charge state.
In this charge state we perform radiofrequency (RF) pulse sequences in resonance with an
NMR transition of the nitrogen nuclear spin.
To ensure a steady charge state a waiting time of 2\,ms is used after the voltage change (see Fig.~\ref{fig:NVplus}d).
Finally, \NVm\ is reset by the proper gate voltage and read out the \Nis\ spin state.
\\
\indent
For identification of \NVp,
we start out with \Nis V centers because they possess a nuclear spin $I=1/2$ and therefore lack quadrupole splitting.
We check for an $m_S=0$ state
in different charge states, by identifying
the corresponding \Nis\ nuclear spin transition $\ket{m_S=0,\, m_I=\downarrow}\leftrightarrow\ket{0, \uparrow}$.
For this purpose, the nuclear spin is initialized into $m_I=\uparrow$ (denoted by $\ket{1_\mathrm{n}}$ in Fig.~\ref{fig:NVplus}c), set a gate voltage and perform an RF $\pi$-pulse in resonance with the mentioned transition to flip the nuclear spin.
As a control experiment, we also track the amplitude of the $\ket{S=1/2,\, m_I=\downarrow}\leftrightarrow\ket{1/2, \uparrow}$ \Nis\ spin transition in the \NVz\ state.
Both nuclear spin transitions are displayed on the left side of Figure~\ref{fig:NVplus}c.
The normalized amplitudes of the latter transitions
equal the probability of being in the corresponding charge state $W_{\pm}$ and $W_0$, where we can not discriminate $W_+$ and $W_-$.
We start out in the \NVm\ state at a gate voltage of $-8\,$V.
$W_{\pm}$ and $W_0$ reveal a switch from \NVm\ to \NVz\ at around $-2\,$V for the NV center investigated (see Fig.~\ref{fig:NVplus}c).
The latter state is stable until around $+8\,$V, where $W_0$ decreases and $W_\pm$ increases again.
Two reasons for this change are conceivable.
Either \NVm\ reappears at higher voltages, or we detect the presence of \NVp\ with an electron spin state $m_S=0$, which is more likely and will turn out to be indeed the case.
Next, we compare nuclear spin Rabi oscillations in the \NVm\ and the \NVp\ case (see Fig.~\ref{fig:NVplus}d).
The smaller Rabi frequency for the \NVp\ case at the same radiofrequency (RF) and the same RF amplitude is explained by the absence of an electron spin and the corresponding impact on the nuclear spin states.
The Rabi frequency ratio is calculated to be 
%\begin{equation}
$\frac{\Omega_{\NVm}}{\Omega_{\NVp}}=1 + \frac{\gammaered}{\gammanred} \frac{2A_\perp D}{\gammaered^2 B_z^2-D^2} =1.832$
%\end{equation}
with the electron and nitrogen nuclear spins' reduced gyromagnetic ratios $\gammaered$ and $\gammanred$, the perpendicular hyperfine interaction $A_\perp$, the Zero-field splitting parameter $D$ and the axial magnetic field $B_z$.
The theoretically derived value agrees well with the experimentally obtained one of $1.81\pm0.04$.
The larger amplitude of the \NVp\ compared to the \NVm\ oscillation amplitude directly relates to the higher probability of being in the newly identified charge state.
The limited spin readout fidelity let's the oscillation in figure \ref{fig:NVplus}d start with a value of 0.3 on the vertical axis and a spin flip with unity probability would yield a maximum value of $1-0.3=0.7$.
Indeed, the displayed oscillations reach 0.7 and hence the chance to be in \NVp\ at $+8\,$V is around one.
However, around $-8\,$V the NV center is only about 70\,\% of time in \NVm\ \cite{waldherr_dark_2011}.
As a result, we have observed around unity combined probability to switch into \NVp\ and preserve the \Nis\ nuclear spin state.
\\
\indent
Having identified the electron spin-less \NVp\ charge state by the \Nis\ probe spin we investigate the accompanying changes of local electric field gradients.
The \Niso\ spin's quadrupole splitting is susceptible to the latter and therefore also is a measure for the change of the charge state (see Fig.~\ref{fig:NVplus}e).
First estimates of the \NVp\ related quadrupole splitting hint to smaller values than for the \NVz\ case.
We therefore employ \emph{ab initio} simulations in order to get insight about the $^{14}$N quadrupole moment of NV defect in diamond as a function of its charge state.
The calculated electric field gradient around the $^{14}$N nuclei in different charge states of NV defect is calculated within Kohn-Sham density functional theory (DFT) that results in qualitatively good agreement with the interpretation of experimental data (see Table~\ref{tab:cqres}, Fig.~\ref{fig:NVplus}e and Methods section for technical details).
In the negative, neutral and positive charge state the NV center is in the $^3A_2$, $^2E$ and $^1A_1$ many-body ground state, respectively, exhibiting $C_{3v}$ symmetry.
Table~\ref{tab:cqres} summarizes our results with NV center in $C_{3v}$ symmetry.
We find that DFT simulations imply a linear decrease in the quadrupole splitting of $^{14}$N when removing electrons from the system.
Our analysis reveals that this effect is caused by the successive depletion of the charge density on the nitrogen atom.
However, experimental data indicate that the $^{14}$N quadrupole moment in NV$^0$ is in magnitude closer to the that in NV$^+$ than to that in NV$^-$ that breaks the linear trend.
We found that this effect is caused by the fact that the $^2E$ ground state is a so-called correlated many-body state that cannot be accurately described by the applied Kohn-Sham DFT method.
We proved by a DFT based configuration interaction method (see Methods for technical details) that accurate calculation of the $^2E$ ground state results in depletion of the charge density on the nitrogen atom with respect to the Kohn-Sham DFT result.
This explains the complex physics and the observed trend on the $^{14}$N quadrupole moment of NV defect with various charge states in diamond.
\begin{table}[h]
	\centering
	\caption{\label{tab:cqres} Experimental (exp.) and \emph{ab initio} calculated (calc.) values for nuclear spin quadrupole splittings of \Niso\ for various NV charge states.
		In the calculations, the $C_q$ quadrupole splitting is $3eQ_{\text{N}}V_{zz}/(4h)$ where $h$ is the Planck-constant and $V_{zz}$ is the corresponding component of the \emph{ab initio} electric field gradient of the potential.
		$Q_{\text{N}}$ is the nuclear quadrupole moment of the $^{14}$N isotope, which scatters between 0.0193 and 0.0208~barn in the literature \cite{stone_table_2005}.
		This uncertainty is reflected in the calculated values.}
	\begin{ruledtabular}
		\begin{tabular}{llll} 
			& NV$^-$ / MHz     &  NV$^0$ / MHz       & NV$^+$ / MHz    \\ \hline %\T %\hline 
			calc.  & -5.02$\pm$0.19   &  -4.92$\pm$0.19     & -4.82$\pm$0.19  \\ 
			exp.   & -4.945           &  -4.655             & -4.619          \\ 
		\end{tabular}
	\end{ruledtabular}
\end{table}
\\
\indent
%
%
%
%
%
%\section{lifetime measurements}
The electron spin-less \NVp\ state is expected to be a formidable storage state for quantum information residing on proximal nuclear spins.
Similar observations were made in ensembles of Si:P
% where the removal of the central electron spin increased the $T_1$ lifetime and the dephasing time of the \Piso\ nuclear spins
\cite{saeedi_room-temperature_2013}.
For the negative charge state \NVm , in contrast, the limiting factor for the nuclear spins' $T_1$ and $T_2$ lifetimes is the hyperfine coupling to the central electron spin.
Whereas the longitudinal relaxation time still reach minutes in moderate magnetic fields ($\sim1\,$T) \cite{pfender_nonvolatile_2016} and therefore set the ultimate limit for $T_2$, the transverse relaxation of the nuclear spins is limited to $T_2\approx10\,$ms by the longitudinal relaxation of the electron spin.
In a spinless environment like diamond, the $T_1$ and $T_2$ values are supposed to increase dramatically by removing the central electron spin.
\\
\indent
We characterize the nuclear spin lifetimes in \NVp\ by preparing the \Niso\ spin in $m_I=1$ ($\ket{1_\mathrm{n}}$) and subsequently setting the appropriate gate voltage (see Fig.~\ref{fig:long_lifetime}b).
The \Niso\ dephasing time is deduced from a spin echo measurement in \NVp\ shown in Figure~\ref{fig:long_lifetime}a.
In addition, a long Rabi oscillation and measurement of the longitudinal relaxation are depicted.
The coherence lifetime is clearly increased.
However, the increase is less than anticipated.
This can be attributed to two effects:
First, the utilized NV centers were created by nitrogen implantation close to the diamond surface, in order for the band bending effect of the gate structure to play a significant role.
Near-surface NV centers commonly suffer from short coherence times on the order of $\sim 10\,\mu$s due to electron spins on the surface and paramagnetic defects created during implantation \cite{favaro_de_oliveira_toward_2016,ofori-okai_spin_2012,romach_spectroscopy_2015,de_oliveira_tailoring_2017} (see Methods section).
Only recently, novel methods have experimentally demonstrated how to overcome these effects (e.g. via plasma treatment of the surface or nano-scale nitrogen doped layers close to the surface) \cite{kim_effect_2014,de_oliveira_effect_2015,ohno_engineering_2012,ohno_three-dimensional_2014,mclellan_patterned_2016}.
If the decoherence is caused by paramagnetic defects on or near the surface, the same effect limits the coherence of the nuclear spin, weakened by the small $g$-factor of the nuclear spin, to the order of $\sim 100\,$ms for a basic spin echo.
Another effect is indicated by the comparably short nuclear spin $T_1$ time of $0.3\,$s.
If the charge state is not stable, other states might induce faster decay.
Indeed, we have observed a very slow and inert response to voltage changes, which we attribute to the large capacitor that enabled a characterization of many NV centers in the first place.
Further experiments with much smaller capacitors will shed light on this behavior.
\\
\indent
%
%
%
%
%
%
%\section{Results and Outlook}
Summarizing we have identified and characterized the positive charge state of NV centers in diamond.
We demonstrated deterministic and reversible electric switching of single NV centers into the newly detected charge state.
It was found that nuclear spin energy eigenstates are resilient under this switching operation and thus serve as probe for \NVp .
They confirmed the absence of an electronic spin, which enabled electron-spin-unlimited nuclear spin coherence storage in \NVp .
Furthermore, the \NVp\ state does not fluoresce under $532\,$nm illumination commonly used for \NVm\ excitation.
Hence, \NVp\ does not contribute to photon shot noise when other proximal NV centers within the same confocal spot are optically interrogated.
\\
\indent
These results pave the way for a fascinating and feasible implementation of the Kane proposal for a scalable, solid-state, spin-based quantum processor \cite{kane_silicon-based_1998,pezzagna_3.13_2014}.
As sketched in Figure~\ref{fig:scalable}a an H-terminated diamond surface would mute all NV based electron spins below 
as already discussed above (see Fig.~\ref{fig:charge_state_switching}).
By adding insulated, nanoscopic top gates selected NV centers might be switched back into \NVm\ upon request.
It is interesting to note, that further increase of the gate potential $V_\mathrm{G}$ would lead to a lateral depletion of the surface conductive channel.
The scale of the lateral depletion effect can be expressed as $r_l \propto V_\mathrm{G}^a/n_s^b$ \cite{denisenko_lateral_2010}, where $n_s$ is the areal charge density in the surface channel and the exponents $a$ and $b$ are roughly in the range of 0.5 and 1.
For H-terminated diamond the $n_s$ value is typically in the $10^{12} - 10^{13}\,\mathrm{cm^{-2}}$ range \cite{maier_origin_2000}.
This allows a precise control of the depletion edge $r_l$ of about few nanometers per Volt of the gate bias \cite{denisenko_lateral_2010}.
This effect would allow individual addressing of closely located NV centers.
Several such NV centers might then coherently interact, for example via magnetic dipole-dipole interaction \cite{dolde_high-fidelity_2014}, and the resulting quantum state can then be stored unharmed on the nuclear storage qubits.
For readout of quantum information, individual NV centers are switched into \NVm , their nuclear spin state is transfered to the electron spin and finally readout optically without touching other nuclear spin qubits of NV centers which reside in \NVp .
\\
\indent
Apart from the potential realization of a scalable quantum processor, the storage of quantum information has proven to be a vital resource for nanoscale quantum metrology.
In Ref.~\onlinecite{zaiser_enhancing_2016} a non-volatile quantum memory enhanced the performance of the NV electron spin sensor.
This enables, on the one hand, coherent interactions of the sensor qubit to spectrally highly selective \Ciso\ spin qubits.
On the other hand, high spectral resolution correlation spectroscopy was demonstrated.
In both cases spectral resolution is inversely proportional to the storage time of the quantum memory.
Our results would yield an increase of spectral resolution by a factor of five.
Furthermore, muting the electron spin sensor also increases the coherence lifetime of sample spins and therefore allow for high resolution spectroscopy in the first place \cite{pfender_nonvolatile_2016}.

\begin{acknowledgments}
We thank Roman Kolesov, Kangwei Xia, Ali Momenzadeh and Jerome Jackson for fruitful discussions and technical advice.
We acknowledge financial support by the German Science Foundation (SFB-TR 21, SFB 716, SPP1601), the Volkswagen Stiftung, the JST and JSPS KAKENHI (No. 26246001 and No. 15H03980), EU grant DIADEMS (grant No.\ 611143).
F. F.O. acknowledges the financial support by CNPq Project No. 204246/2013-0.
\end{acknowledgments}

\section{Methods}
\footnotesize
\textbf{The experimental setup} consists a home-built confocal microscope with an oil immersion objective at room-temperature.
Furthermore, a permanent magnet creates a field of $B_z\approx 470\,$mT along the optical axis, which is collinear with the NV center's symmetry axis.
Microwave (MW) and Radiofrequency (RF) fields for electron and nuclear spin gate realizations are provided through a copper wire close to the NV center's position.
\\
\indent
\textbf{Diamond sample preparation.}
(111)-oriented diamond plates were obtained by laser-slicing of a low-strain type-IIa HPHT crystal.
One of them was polished and used as a substrate.
High-purity homoepitaxial diamond (111) film was deposited using a microwave plasma-assisted chemical-vapor-deposition (MPCVD) apparatus \cite{teraji_homoepitaxial_2015}.
High-purity H$_2$, CH$_4$ specified to 99.999\% \Cis\ isotopic enrichment (Cambridge Isotope Laboratories CLM-392), and high-purity O$_2$ was used as source gases.
The total gas pressure, microwave power, methane concentration ratio (CH$_4$/total gas), oxygen concentration ratio (O$_2$/total gas), growth duration and substrate temperature employed were $120\,$Torr, $1.4\,$kW, 1\%, 0.2\%, $11\,$h, and $950-1000^{\circ}$C, respectively.
Growth rate under this condition is $3.3\,\mu$m/h, thus the homoepitaxial film thickness is deduced to be $\approx 36\,\mu$m.
The pristine MPCVD layer contained a negligible amount of NV centers.
We implanted $^{15}\mathrm{N}^{+}$ and $\mathrm{He}_2^{+}$ ions with kinetic energies of $10\,$keV and $6\,$keV and doses of $8\cdot10^9\,\mathrm{cm}^{-2}$ and $10^{10}\,\mathrm{cm}^{-2}$ for creation of sub-surface NV centers \cite{pezzagna_creation_2010,favaro_de_oliveira_toward_2016}.
While the \Nis\ implantation mainly leads to \Nis V centers, helium creates vacancies that form \Niso V centers from ingrown \Niso\ impurities.
The expected average depths of the created NV centers are $25\,$nm \cite{de_oliveira_effect_2015}
and $10-30\,$nm \cite{favaro_de_oliveira_toward_2016} for the nitrogen and the helium implantations, respectively.
For final NV center formation the sample was annealed at $950^{\circ}$C for two hours.
The sample was then boiled in acid to remove graphite.
\\
\textbf{In-plane gate electrodes}
were created as described in \cite{hauf_addressing_2014}.
Hydrogen plasma treatment of the surface creates a conductive surface layer, which switches NV centers into the \NVp\ state.
Interdigitated capacitors are created by electron-beam lithography and substituting hydrogen by oxygen surface terminations in the selected regions (see Fig.~\ref{fig:charge_state_switching}c).
The new termination in the gaps of the capacitor is nominally not conductive and has a lateral width of $\approx 500\,$nm.
The large capacitor with an insulating gap length of more than $1\,$mm (see fig.~\ref{fig:charge_state_switching}) allows for investigation of many NV centers.
Figure~\ref{fig:NVplus}a reveals a sheet resistance on the order of $10^{11}\,\Omega/\text{m}^2$, which leads to a finite current for a few volts of bias field.
The capacitor electrodes were fabricated on top of those regions containing NV centers from ion implantation.
The electrodes of the capacitor are connected to large H-terminated areas, where gold pads are evaporated and connected by gold wire bonding to the periphery.
\\
\textbf{The Spin Hamiltonian}
restricts to the NV center electron spin (S=1) and \Nis , \Niso\ nuclear spin ($I=1/2,\,I=1$) for the \NVm\ case (eq.~\ref{eq:H_tot}).
For the \NVp\ case only the pure nuclear spin parts ($H^\mathrm{n}$) remain including the corresponding quandrupole splitting $Q_{-} \rightarrow Q_{+}$.
The electron ($H^\mathrm{e}$) as well as hyperfine coupling ($H^\mathrm{hf}$), however, can be neglected.
\begin{align}
	\label{eq:H_tot}
	%H =& H^\mathrm{e} + H^\mathrm{n} + H^\mathrm{hf} \\
	  %=& DS_z^2 + \tilde{\gamma}_\mathrm{e} B_z S_z + Q_{-}I_z^2 + \tilde{\gamma}_\mathrm{n} B_z I_z \nonumber \\
		%&+ A_\parallel S_z I_z \nonumber + A_\perp/2 \left( S_{+}I_{-} + S_{-}I_{+} \right)
	H =& \overbrace{DS_z^2 + \gammaered B_z S_z}^{H^\mathrm{e}} + \overbrace{Q_{-}I_z^2 + \gammanred B_z I_z}^{H^\mathrm{n}} \\
		&+ \underbrace{A_\parallel S_z I_z \nonumber + A_\perp/2 \left( S_{+}I_{-} + S_{-}I_{+} \right)}_{H^\mathrm{hf}}
\end{align}
In eq.~(\ref{eq:H_tot}) $D$ is the electron spin triplet zero-field splitting, $\gammaered$ and $\gammanred$ are the electron and nuclear spin gyromagnetic ratios over $2\pi$, and $A_\parallel$ and $A_\perp$ are the longitudinal and transverse hyperfine coupling constants to the nitrogen nuclear spin.
When an electron spin is present, nuclear spin states get slightly dressed and thus deviate from pure product states due to hyperfine interaction.
This effect is observable in nuclear spin Rabi oscillations, which get faster or slower than for a pure nuclear spin.
A first order perturbation of the product states is sufficient for an estimation.
For a \Nis V center the $m_S=0$ spin levels change according to
$\ket{0\uparrow} \mapsto \ket{0\uparrow} - \frac{A_\perp}{\sqrt{2}\left(\gammaered B_z + D\right)}\ket{+1\downarrow}$ and
$\ket{0\downarrow} \mapsto \ket{0\downarrow} + \frac{A_\perp}{\sqrt{2}\left(\gammaered B_z - D\right)}\ket{-1\uparrow}$,
where $\uparrow,\downarrow$ denotes the state of the \Nis\ spin.
We derive a \Nis\ nuclear spin Rabi frequency enhancement factor of $1 + \frac{\gammaered}{\gammanred} \frac{2A_\perp D}{\gammaered^2 B_z^2-D^2} =1.832$ at $470\,$mT and for $A_\perp=3.689\,$MHz.
The latter value is derived from the \Niso\ value of $A_\perp=-2.630(2)\,$MHz by taking into account the nuclei's different $g$-factors.
\indent
\\
%\paragraph{Methods}
\textbf{Density functional theory based calculations}
The density functional theory (DFT) simulations were carried out within plane wave supercell formalism together with the all-electron projector augmented wave (PAW) method \cite{blochl_projector_1994} as implemented in \textsc{vasp} scientific code \cite{kresse_textitab_1993,kresse_textitab_1994,kresse_efficiency_1996,kresse_efficient_1996}.
The electron-nuclei system is treated within Born-Oppenheimer approximation where the quantum mechanical problem of electrons and the nuclei systems is separated, and the nuclei are approximated as classical particles.
We show the results in the main text as obtained by Perdew-Burke-Ernzerhof (PBE) DFT functional \cite{PBE_1996}.
We found that other DFT functionals including hybrid density functionals provided the same trend for the quadrupole moment of $^{14}$N in various charge states of the NV defect.
The NV defect was modelled in a 512-atom simple cubic diamond supercell with PBE optimized lattice constant.
The Brillouin zone is sampled in the $\Gamma$-point which is convergent within 1\%.
A high plane-wave cutoff of 1200~eV is applied.
The geometry of the NV center was optimized in each charge state until the forces on the atoms fell below 0.0001~eV/\AA .
In the calculation of the electric field gradient on nitrogen nuclei we applied a hard PAW-potential for nitrogen as provided by \textsc{vasp} package. Since NV$^0$ is a dynamic Jahn-Teller system we calculated its potential energy surface around the $C_{3v}$ configuration. We found that the $C_{1h}$ configuration has the lowest energy within Born-Oppenheimer basis. The absolute value of the calculated nuclear quadrupole moment of $^{14}$N is reduced by about 3.7\% with respect to that of $C_{3v}$ configuration. By analyzing the DFT orbitals in the reduced $C_{3v} \rightarrow C_{1h}$ symmetry we found that the most important effect on the calculated nuclear quadrupole moment of $^{14}$N is the correlation between the $a_1 \rightarrow a'$ and $e \rightarrow a', a''$ Kohn-Sham DFT orbitals in the band gap. The highly correlated localized orbitals cannot be accurately captured by Kohn-Sham DFT functionals. Indeed, the $^2E$ ground state within $C_{3v}$ symmetry is a multiplet state of $a_1^{(2)}e^{(1)}$ and $a_1^{(1)}e^{(2)}$ electron configurations (where the numbers in parentheses label the occupation of 
orbitals). In order to calculate these many-body $^2E$ states, we applied an approximate configuration interaction based on DFT wavefunctions \cite{wang_linear_1999,franceschetti_many-body_1999,bester_electronic_2009}. We found a 5\% 
contribution of $a_1^{(1)}e^{(2)}$ configuration to the $^2E$ ground state. Since only the $a_1$ orbital is localized on the nitrogen atom \cite{gali_ab_2008} this causes some charge density delocalization around the nitrogen atom which consequently decreases the absolute value of the calculated nuclear quadrupole moment of $^{14}$N with respect to that of DFT values.

\normalsize
%\bibliographystyle{merlin3}
%\bibliography{RefsZotero,bib}
%\input{nvplusnn.bbl}

\end{document}